**Building Workflows for Interactive Human in the Loop Automated Experiment (hAE) in STEM-EELS**

Utkarsh Pratiush[1], Kevin M. Roccapriore[2], Yongtao Liu[2], Gerd Duscher[1], Maxim Ziatdinov[3], and Sergei V. Kalinin[1,3]

[1] Department of Materials Science and Engineering, University of Tennessee, Knoxville, TN 37996, USA

[2] Center for Nanophase Materials Sciences, Oak Ridge National Laboratory, Oak Ridge, TN 37831, USA

[3] Pacific Northwest National Laboratory, Richland, WA 99354

Exploring the structural, chemical, and physical properties of matter on the nano- and atomic scales has become possible with the recent advances in aberration-corrected electron energy-loss spectroscopy (EELS) in scanning transmission electron microscopy (STEM). However, the current paradigm of STEM-EELS relies on the classical rectangular grid sampling, in which all surface regions are assumed to be of equal *a priori* interest. This is typically not the case for real-world scenarios, where phenomena of interest are concentrated in a small number of spatial locations. One of foundational problems is the discovery of nanometer- or atomic scale structures having specific signatures in EELS spectra. Here we systematically explore the hyperparameters controlling deep kernel learning (DKL) discovery workflows for STEM-EELS and identify the role of the local structural descriptors and acquisition functions on the experiment progression. In agreement with actual experiment, we observe that for certain parameter combinations the experiment path can be trapped in the local minima. We demonstrate the approaches for monitoring automated experiment in the real and feature space of the system and monitor knowledge acquisition of the DKL model. Based on these, we construct intervention strategies, thus defining human-in the loop automated experiment (hAE). This approach can be further extended to other techniques including 4D STEM and other forms of spectroscopic imaging. The hAE library is available at (Github: https://github.com/utkarshp1161/hAE/tree/main/hAE)

Electron energy-loss spectroscopy (EELS) in scanning transmission electron microscopy (STEM)[1] has emerged as a transformative technique in modern materials science, offering an unparalleled window into the structural and electronic properties[2,3] of materials at the atomic and nano scales. This tool has been crucial in the development of advanced nanomaterials[4], semiconductor technology, and nanoelectronics, enabling breakthroughs in these fields. It has also contributed significantly to energy research, particularly in solar cells[5] and battery materials[6], and in analyzing and optimizing catalysts[7] for more efficient chemical processes. Furthermore, STEM-EELS has provided critical insights into plasmons in nano-optical structures and the study of quasiparticles and vibrational excitations[89], advancing fundamental physical understanding in fields like photovoltaics, sensing technologies, and solid-state physics.

The current paradigm of STEM-EELS is based either on single point spectroscopic measurements or hyperspectral imaging on rectangular grids. For the former, human operator selects the measurement locations based on the structural features observed in the structural STEM image. It is also important to note that in many cases the imperfection in the scanning systems can result in the misalignment between the intended and actual measurement points, leading to difficult to detect errors. For the second, the region of interest is identified, and multiple EELS spectra are acquired over rectangular grid of points. The resulting 3D hyperspectral data set can be analyzed using physics-based methods[10] or linear[11] or non-linear[12] dimensionality reduction methods to yield 2D images that are amenable to human perception and potentially interpretation. However, in this approach the information is uniform over the image plane, whereas in most materials systems the objects of interest are typically localized in a small number of locations. Similarly, this approach is typically associated with significant beam damage and large acquisition times.

The limitations of the classical EELS measurements and recent emergence of the Python interfaces to commercial instruments have resulted in strong interest towards automated spectroscopic measurements. One such approach is based on combination of the application of computer vision-based approaches to identify a priori known objects of interest and subsequent spectroscopic measurements on these chosen locations.[13–16]

The alternative approach to automated EELS measurements is represented by the inverse workflow.[17] In this, the spectral signature of interest such as specific peak positions, integrated intensity, peak ratios, etc. is identified based on the prior knowledge and intended goals of the experiment. This scalar measure of physical interest is referred to as the scalarizer function, and the purpose of the experiment is to discover microstructural elements at which this scalarizer is maximized. In this sense, DKL is targeting the exploration of physics of interest as reflected in a spectrum[18]. The example of such an approach is DKL based workflows recently demonstrated by Roccapriore for EELS and 4D STEM[19,20], and Liu for several SPM modalities[21,22]. There is also a study which times the performance gain of the DKL in capturing the spectrum-imaging which is speed's up more than 300x[23].

However, until now the DKL[24] AE workflows[14,25] were realized using largely ad hoc hyperparameter values chosen before the experiment. These included the choice of scalarizer and acquisition functions defining exploration-exploitation balance during experiment. Over the last 2 years, we have observed that the experimental path can be strongly affected by these parameters, sometimes resulting in the process being stuck at selected locations or exploring only one specific type of microstructural elements. This sensitivity to hyperparameters and propensity to be trapped in metastable minima is well known for ML methods.[26–28] However, for active learning problems on the experimental tools, the classical strategies for hyperparameter tuning are inapplicable. While some parameter optimization can be achieved using pre-acquired ground truth data, this approach is sensitive to out of distribution shift effects even for similar samples and cannot be expected to generalize for different materials.

Correspondingly, implementation of the AE for the STEM-EELS experiments requires the introduction of a different paradigm based on the interactive, or human in the loop, hAE. In this hAE approach, the human operator monitors the progression of the AI-driven automated experiment and introduces high-level modifications in the policies that govern the actions of the machine learning agent at each step of the experiment. This integrative approach between AI and human was proposed for SPM(Scanning probe microscopy);[25] however, the nature of the possible control parameters, exploration policies, and their effects on the exploration pathway has been unexplored.

Here, we present a benchmark study across a comprehensive range of hyperparameters, including local structural descriptors and various acquisition functions (AFs) tailored for both exploitation and exploration phases. Additionally, we explore different AF parameters to optimize our experimental setup. A pivotal aspect of our methodology involves monitoring the learning progression within both the real and feature spaces of the system. The latter can in turn be defined via the Variational Autoencoder (VAE[29–31]) approach. This dual monitoring provides critical insights into the progression of automated experiments. Through a detailed study, we quantify these observations, establishing a direct link to the pivotal role humans play in selecting the appropriate parameters. This discussion further extends to potential human interventions, highlighting the balance between automated processes and human expertise in optimizing experimental outcomes. Our findings underscore the significance of human intuition and decision-making in refining and guiding automated experimental workflows.

## I. General setting of automated experiment

The general setting of the STEM-EELS experiments is illustrated in Figure 1 and can be generalized for any imaging experiments based on structural and spectral images. Structural data **S(**$x$,$y$**)** are easy to acquire and have high density in image plane, but have relatively low information density per pixel in one or few channels. In comparison, spectroscopic data **A(**$E$**)**, Figure 1b, contains a wealth of information on materials properties, but acquiring spectroscopic data is time consuming. Thus, by now standard approach is to acquire both the structural and the hyperspectral data sets **A(**$x$,$y$,$E$**)** , Figure 1c, over rectangular grid of points. For hyperspectral data and analyze using ML or physis based methods to get a set of 2D maps.

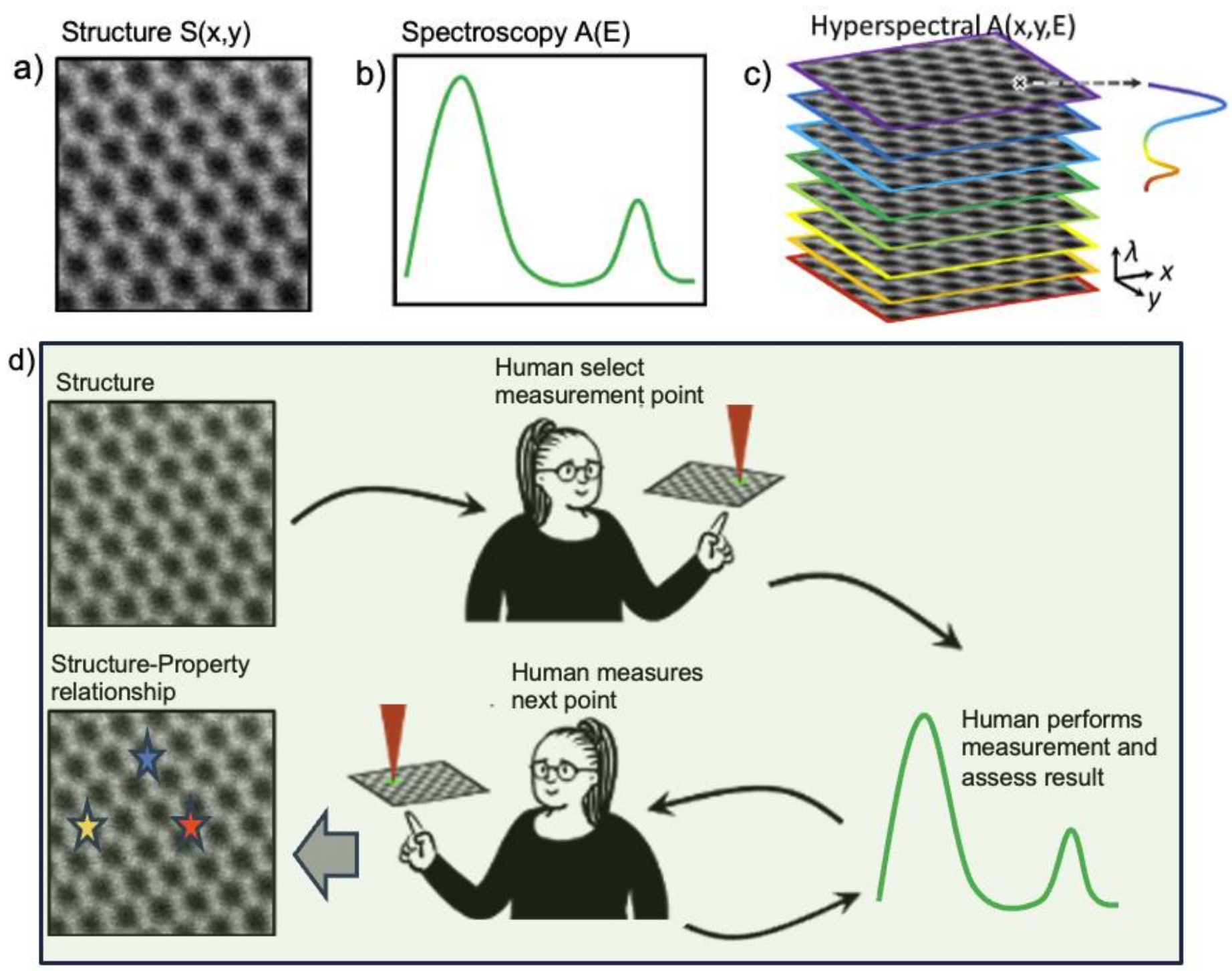

**Figure 1.** An example of (a) structural and (b) spectral data (coming from a single point) available in STEM. Generally, structural images are fast to acquire, whereas spectral acquisition is more time consuming. In (c) hyperspectral imaging, the spectra are acquired over dense grid, giving rise to 3D (or higher-dimensional) data set. Alternatively, (d) human selects the measurement locations based on structural elements of interest, and iteratively explores image plane collecting spectral data. Overall, the goal of the experiment is to explore physical behaviors of interest visible in the spectral data using the structural images as a guide. Note that hyperspectral imaging acquires spectral data everywhere, whereas human (or ML based) feature identifications identifies location of interest based on the features visible in structural image (direct workflow). The inverse workflow as realized in DKL(as shown in Figure 3) solves inverse problem – discovery of structural features based on their spectral signatures of interest.

The rectangular sampling is easy to implement and, after suitable dimensionality reduction[32], is readily amenable to human perception. It is also optimal in a sense that natural way to sample unknow space if we have no prior information. However, in most cases the information of interest is concentrated in specific spatial locations. The grid-based measurements in this case are sub-optimal and are associated with potential for beam damage and small explorable areas because large grids result in long acquisition times longer than microscope stability allows.

The further development are dynamic techniques that use prior information to create sampling pattern. For spectral methods, the natural approach is to use structural information to identify objects of interest. One way to do it is via human selection to identify objects of interest, i.e. create the locations for the spectral measurements. A similar paradigm can be implemented via deep convolutional neural network (DCNN) trained on human-labeled data[33] or discovering elements in the unsupervised manner[34]. Once the data is acquired, it can be analyzed to build structure-property relationships, and if necessary, expanded to yield 2D images via variants of inpainting, Gaussian processing[35], etc. However, this strategy relies on *a priori* knowledge of which structural objects comprise the information of interest or purely on statistical properties of objects in the image plane.

At the same time, in many scenarios it is the specific aspects of **spectral** data that we aim to discover. The examples include signatures of quasiparticles, valence states, signatures of surface and bulk plasmons, peaks or edge onsets corresponding to specific elements, peak ratios related to the oxidation states or orbital populations, and so on.[36–38] In these cases, we can introduce the measure of physical interest $A(E)$ that maps the EELS spectrum to a signature of interest. The scalarizer $P$ can be a scalar functional of spectra, or a (low dimensional) vector. Hence, a setting for automated experiment is whether we aim to explore the image plane $I(x,y)$ based on the features of interest in the spectra. For example, this can be learning the relationship between the structural features and spectra, or discovery of structural features that give rise to certain signatures, i.e., those that maximize $P$. For scalar $P$ this is optimization, for vector $\mathbf{P}$ multiobjective optimization.

Overall, there are three primary strategies for the spectroscopic experiments. The classical approach is to acquire $\mathbf{A}(x,y,E)$ based on structural domain using a rectangular grid, shown in Figure 1c. This approach is slow and associated with the beam damage to the sample. This strategy is optimal when we have no other information regarding the sample (or if the structural and spectral data are uncorrelated). The alternative approach is to acquire $A(E)$ manually based on structural domain and supervision of human at each step, as shown in Figure 1d. This method can be expensive and heavily biased by the level of expertise of the human, and strongly affected by the imperfection of the positioning system. Finally, the third approach is the inverse approach. Here, the goal of the experiment is to sequentially acquire EELS spectra to discover at which structural elements in real space does the certain spectrum manifest. To achieve this goal, the ML algorithm learns a mapping between structure domain and the spectrum domain.

The established approach for inverse experiment workflows is based on Deep Kernel Learning[18] (DKL). To illustrate mathematical foundations of the DKL, we first consider the simple Gaussian Process, defined as:

$$f(x) \sim \mathcal{GP}\big(m(x), k(x, x')\big) \tag{1}$$

In Eq.(1), $\mathcal{GP}(x)$ represents the Gaussian Process, where $x$ is the parameter space. The $f(x)$ is the function we observe (e.g. image contrast). The $m(x)$ is the mean function of the GP, which describes the expected value of the process at input. In typical imaging problems $m$ is taken as zero. Finally, $k(x, x')$ is the covariance (kernel) function of the GP, which models the dependencies between different input points x and x'. Basically, GP represents the strategy to interpolate unknown black-box function, yielding the surrogate model that predicts the function value and its uncertainty over the full parameter space. These predictions and uncertainty can be further used to guide the active learning over this parameter space, i.e. guide the selection of the next measurement points. Bayesian Optimization is an example of such an approach, as will be discussed later.

Conversely, deep kernel learning learns the representation of some unknown function from some high dimensional descriptor, building the correlative relationship between the two. In the context of the STEM-EELS experiment, the high dimensional descriptor can be chosen to be the local structure within a certain sampling window (image patch), whereas discoverable function is either the full EEL spectrum or some representations of the spectrum (scalarizer).

A deep kernel learning[24] is defined as:

$$k_{DKL}\big(x_i, x_j \big| w, \theta\big) = k_{base}\big(g(x_i|w), g\big(x_j\big|w\big)\big|\theta\big) \tag{2}$$

In Eq.(2), $g$ represents a neural network characterized by its weights $w$, while $k_{base}$ denotes a standard Gaussian process (GP) kernel (e.g. RBF or Matern)[39]. The neural network's parameters and those of the GP base kernel are jointly learned through either Markov chain Monte Carlo (MCMC) sampling methods or stochastic variational inference. Following training, the resulting Deep Kernel Learning (DKL) model is employed to acquire predictive mean and uncertainty values, as well as to construct the acquisition function, similar to standard GP. The difference between the two is the standard GP learns to reconstruct the black-box function by iterative sampling (e.g. reconstruct the image), whereas the DKL actively learns the correlative relationship between known structural and dynamically explored spectral data (i.e. learn structure-property relationships).

## II. The DKL automated experiment

The steps of the DKL automated experiment (AE) include selection of scalarizer and control of the exploration-exploitation balance via acquisition function. These elements and the outputs of the AE are discussed in detail below using the pre-acquired STEM-EELS data set on fluorine and tin co-doped indium oxide infrared plasmonic nanoparticles. A monochromated electron beam (~50 meV full width half max) was used to access the near infrared spectral regime where the plasmon resonances of these nanoparticles exist. Other relevant conditions and sample preparation are reported in elsewhere;[40–42] the analysis here was performed on the data sets obtained under equivalent conditions.

### II.1. Selection of scalarizer:

By our definition, the scalarizer function defines the measure of scientist's interest to a spectrum *A*(E). In the active learning terminology, scalarizer is the myopic (i.e. available at each experiment step) reward function. Scalarizer is designed with the help of the domain scientist based on knowledge of the material and enabling scientists to explore the material's properties with a high degree of specificity and relevance to the domain of interest. Here, we discuss the possible definitions of the scalarizer function for the EELS on nanoplasmonic particles.

Here, the electron beam excites multiple plasmon modes within the nanoparticle cluster in Figure 2, which exist at different locations in space and different energies. Three primary plasmon modes are shown in Figure 2: a low energy and long-range collective mode (dipole mode), a mode confined to the particle edges (edge mode), and a mode confined to nanoparticle interior (bulk mode). Since these modes occur at different energies, they can be selectively imaged from a hyperspectral image by integrating the energy band associated with each plasmon mode. Correspondingly, we can define a scalarizer as a spectral bandpass filter: the scalarizer 1 captures dipole mode, 2 captures edge mode, 3 captures bulk mode. Dependent on the experiment goals, the scalarizer function can be chosen to be more complex – e.g. peak height ratio, peak width, asymmetry, and any other functional of the spectrum.

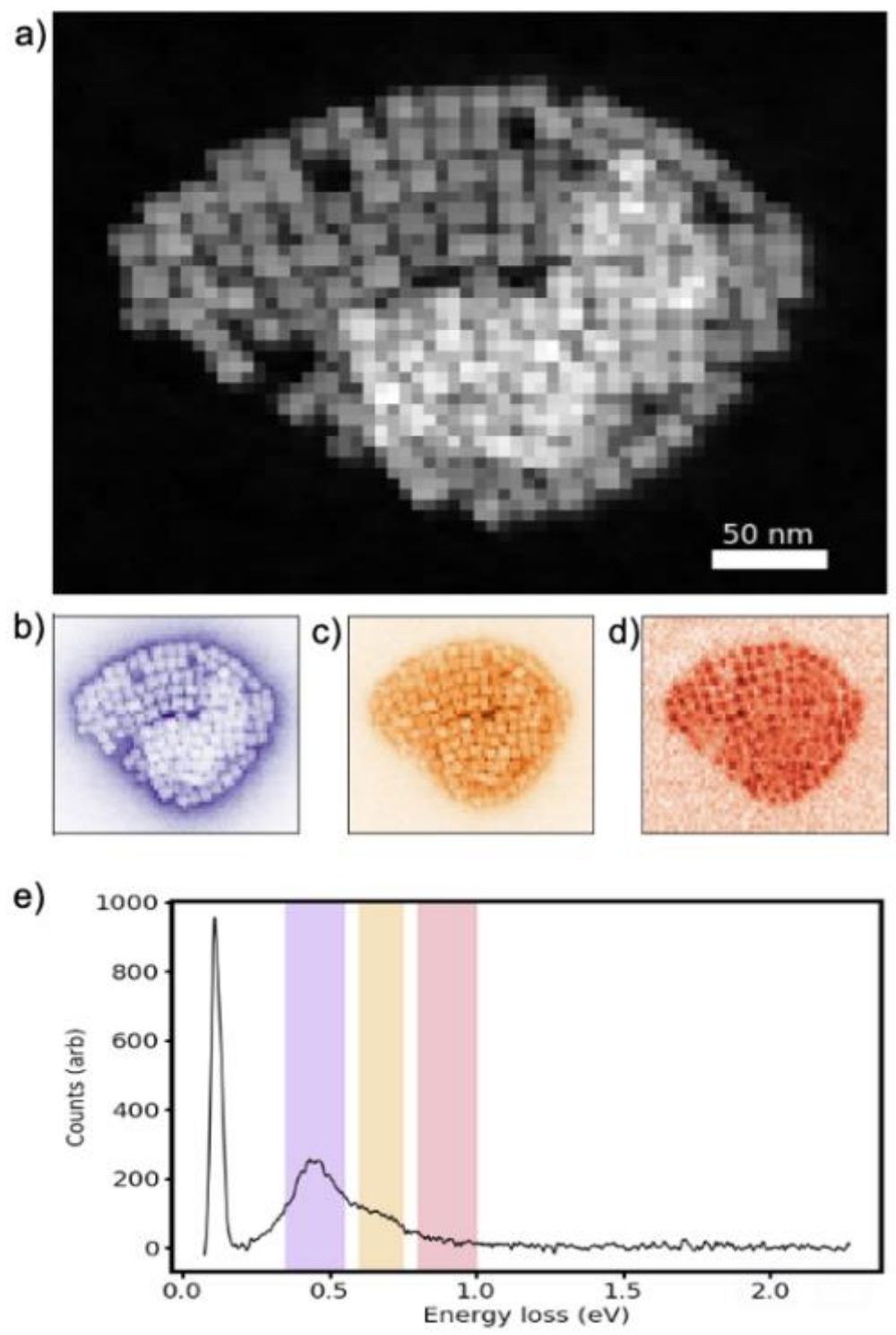

**Figure** 2. This figure represents the nanoplasmonic system: (a) is the HAADF image detailing nanoparticle structure; (b), (c), and (d) are scalarizer images representing the dipole, edge, and bulk plasmon modes, highlighting different plasmonic oscillations; and (e) is the averaged EELS spectrum, with scalarizer regions exemplified. **Note we use the terminology scalarizer 1, 2 and 3 for dipole, edge and bulk mode respectively in the paper.**

It is important to note that previously scalarizer was defined before the experiment. However, in the interactive hAE the scalarizer can be dynamically tuned during the experiment, or the type of the scalarizer can be changed. For example, the boundaries of the region of integration can be dynamically adjusted, or more complex scalarizers such as peak fit parameters, peak ratios, etc. can be chosen. These behaviors will be explored in the sections below.

### II.2. Policies in DKL

Policy refers to a strategic approach that guides the decision process of the ML algorithm. For the myopic optimization frameworks, the policy is generally built based on the estimated prediction and uncertainty of prediction for unexplored parts of the parameter space, where predictive model is built based on prior measurements via GP or DKL. In the context of the STEM-EELS experiments, policy determines the selection of locations for new EELS measurements based on structural image and results of previous EELS measurements.

The DKL policy is determined by the acquisition function parameters which is tuned for exploration and exploitation. While the number of possible policies is large and new policies can be formulated, the basic policies include expected improvement (EI), upper confidence bound (UCB), and mean utility. These acquisition functions[43] are explained below.

$$EI(x) = \int_{-\infty}^{\frac{(f(x_{best}) - \mu(x))}{\sigma(x)}} \left(f(x) - f(x_{best})\right) \cdot \varphi(z) dz \quad (3)$$

Where x is the point to evaluate, $\mu(x)$ is the predictive mean, $\sigma(x)$ is the predictive standard deviation, f(x) the function value at (x), $f(x_{best})$ is the current best observed function value, $\phi(z)$ is the standard normal probability density function, and z is a standard normal random variable.

The expected improvement (**EI**) acquisition function (equation 3) is typically used to balance exploration and exploitation. It encourages the selection of points where the predicted function value is likely to improve upon the current best value. Expected Improvement (EI), measures the expected improvement over the current best observation. However, in some scenarios, one might want to explicitly maximize the model's uncertainty (**MU**) to improve exploration, especially in the early stages of the search or when the global structure of the function is unknown. This can help to ensure a more thorough search of the parameter space and avoid premature convergence to local optima.

Finally, the Upper Confidence Bound acquisition function (equation 4) balances exploration and exploitation by selecting points based on both the predictive mean and the predictive standard deviation (uncertainty) of the GP model.

$$UCB(x) = \mu(x) + \delta \cdot \sigma(x) \quad (4)$$

Where $\mu(x)$ is the predictive mean of the GP at point (x), $\sigma(x)$ is the predictive standard deviation (uncertainty) of the GP at point (x), and $\delta$ is the UCB parameter that determines the balance between **exploration and exploitation.**

In realistic settings, the acquisition function can also include the cost of measurements. This cost can be either *a priori* known, or be a discoverable function of the experiment, e.g. predicted over the full image space by a separate Gaussian Process. However, in STEM EELS we assume that the measurement costs are equal for all locations within the image plane.

Similarly to the scalarizer, the acquisition function has control parameters, e.g. exploration – exploitation balance. During the automated experiment we can consider dynamically tuning the acquisition function or switching between different acquisition functions. We also note that spectrum acquisition can be driven by a different strategy, e.g. random selection of points or sampling specific structural features. These can be switched dynamically during the experiment. Like the scalarizer, this requires approaches to monitor the AE to make these decisions.

### II.3. Experiment progression and output.

With the scalarizer and acquisition functions defined, we discuss the general setting of the DKL experiment. The goal of the DKL experiment is to discover which structural features in the $S(x,y)$ maximize $P = P(A(E))$. To accomplish this goal, the image is represented as a collection of M patches each of size NxN, where N is the patch size and M is the total number of patches. Each patch scan be indexed by location $(x_i,y_i)$, where $x_i,y_i$ correspond to the point on the global image $S(x,y)$ from which the patch has been taken. In other words, the patches are sampled over the rectangular grid. All the patches are available from the beginning.

The microscope performs the AE on a set of seed patches (can be a single) to generate a set of spectra $\mathbf{A_i}(xi,yi,E)$ and hence evaluate the scalarizer in these locations, $P(x_i,y_i)$. The seed points $x_i$, $y_i$ can be chosen randomly, or selected based on the analysis of the features in the global image $S(x,y)$. The DKL algorithm is trained on all the patches $s(x_j,y_j)$ and the scalarizer functions available in the locations $(x_j,y_j)$.

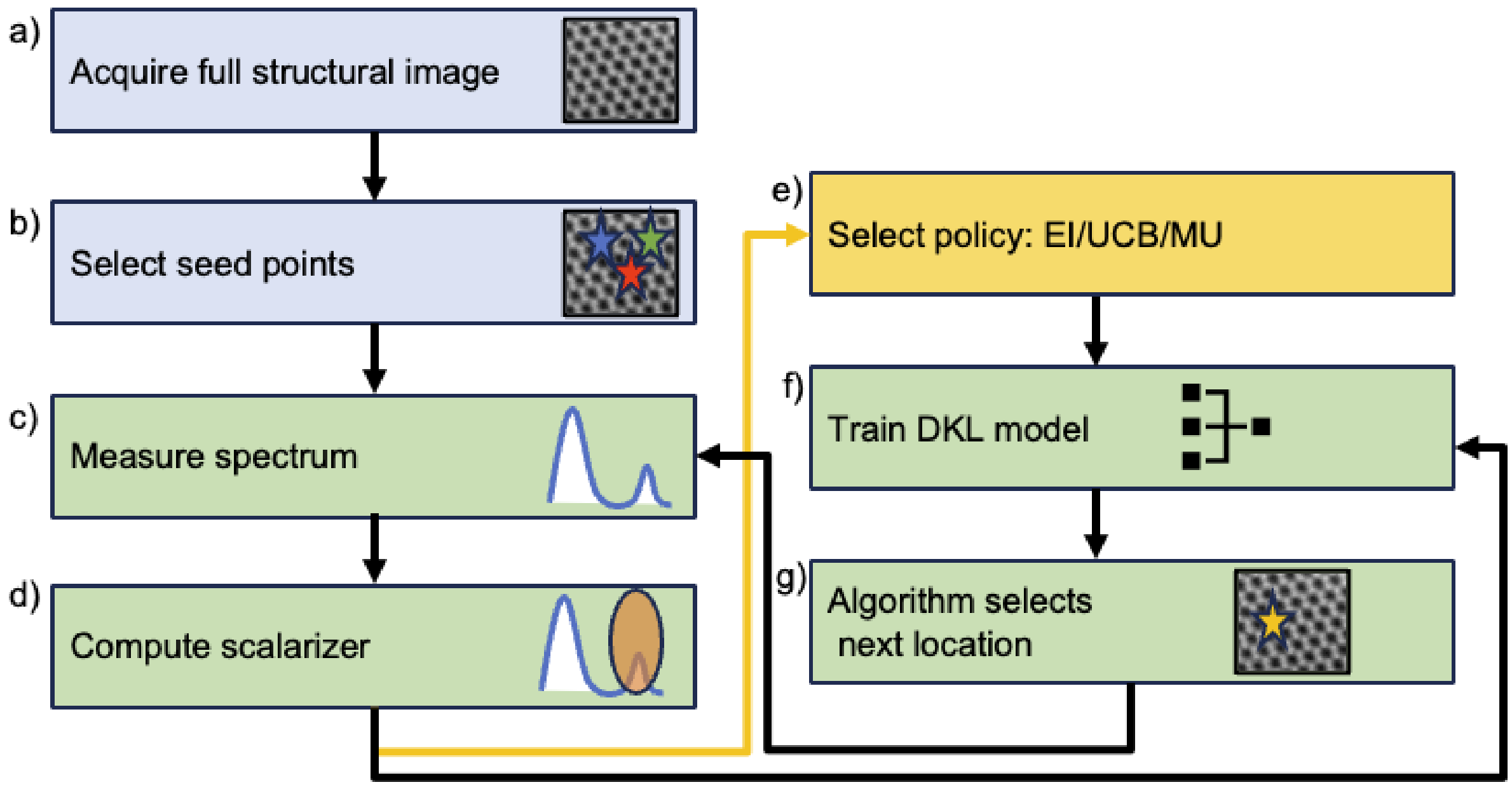

**Figure 3.** Deep kernel learning workflow: a) Acquire an image and extract patches around the atoms. b) Select seed points c) Perform spectroscopy at the seed points, generating patch-scalarizer pairs, d) Select active learning policy, f) Train deep kernel model g) Select next point based on policy. Steps c)-d)-f)-g) repeat until experimental budget. Notice e) that is policy, is set once at start of experiment.

After training and prediction, the DKL algorithm performs the spectral measurement in the patch with coordinates:

$$(x_n, y_n) = \text{argmax}\,(\text{Acq}(x_i, y_i)) \tag{5}$$

Where $i$ goes over all $M$ patches (i.e. all structural descriptors). With the new measured spectrum $\mathbf{A_n}(x_n, y_n, E)$, the scalarizer function, $P(A_n)$ is calculated and the DKL algorithm is retrained with the additional data set. This is repeated until the expiration of experimental budget. Finally, experimental trace is defined as the sequence of patches, locations, and measured spectra acquired during the experiment progression.

### II.4. Additional Types of DKL experiment

The output of the DKL AE is the experimental trace, or collection of the patches and corresponding spectra. With the trace and after the experiment, we found it useful to define several forensic tools[44] that make the understanding of the DKL easier. These include:

The "DKL Explore" process is designed to systematically explore a dataset using Deep Kernel Learning (DKL) techniques. It begins by preparing the dataset, extracting patches and associated scalarizer values, and splitting the data into training and testing sets. Over a series of exploration steps, a DKL model is trained on the training data, enabling the prediction of scalarizer mean and variance values for all data patches. The selection of the next data point is guided by an acquisition function, which aids in identifying valuable information. The chosen data point, along with its associated scalarizer value, is then added to the training data. This process is repeated for each step, recording critical information such as mean, variance, selected index, acquisition function value, and scalarizer value. Finally, the final training and testing datasets are saved. This approach allows to simulate the DKL over pre-acquired data set, and do the initial parameter tuning, reveal the relationships between the parameters, etc.

The "DKL Counterfactual" process conducts dataset exploration with a unique focus on counterfactual scenarios within the context of Deep Kernel Learning (DKL). It initiates by collecting data patches and their associated scalarizer values and establishes an initial train-test split. Over each exploration step, a DKL model is trained on the existing training data to facilitate the prediction of scalarizer mean and variance values for all data patches. This process employs records from previous exploration steps to inform the selection of the next data point, without relying on traditional acquisition functionssl This counterfactual approach allows for a comprehensive examination of alternative scenarios and a deeper understanding of the automated experiment trajectory.

We have further summarized related terminology **live, final** and **complete** model required to monitor knowledge acquisition in the DKL experiment in Table 1. The table also contains definition to terminologies, **Latent space**, **Latent embedding** and **Latent images** which are introduced later in the paper

## III. DKL on full data and the role of the window size

The DKL experiment is defined in a large space of hyperparameters corresponding to the selection of patch (window) sizes, scalarizer function, acquisition function, and their hyperparameters. Hence, similarly to classical ML, it is advantageous to examine the effects of these hyperparameters using the pre-acquired data.

As a first step, we have explored the effect of the varying patch sizes. We have systematically explored 5,10 and 15 window sizes. Each patch is measured in terms of there area whose units are pixel square. We have noted that increasing the patch size results in the change in the effective resolution in the latent embedding image as shown in Figure 4[**see also Supplementary for quantification**]. We observed a distinct scalarizer pattern emerging in the DKL embedding, which indicates that DKL is effectively learning the structure-property relationship inherent within the data.

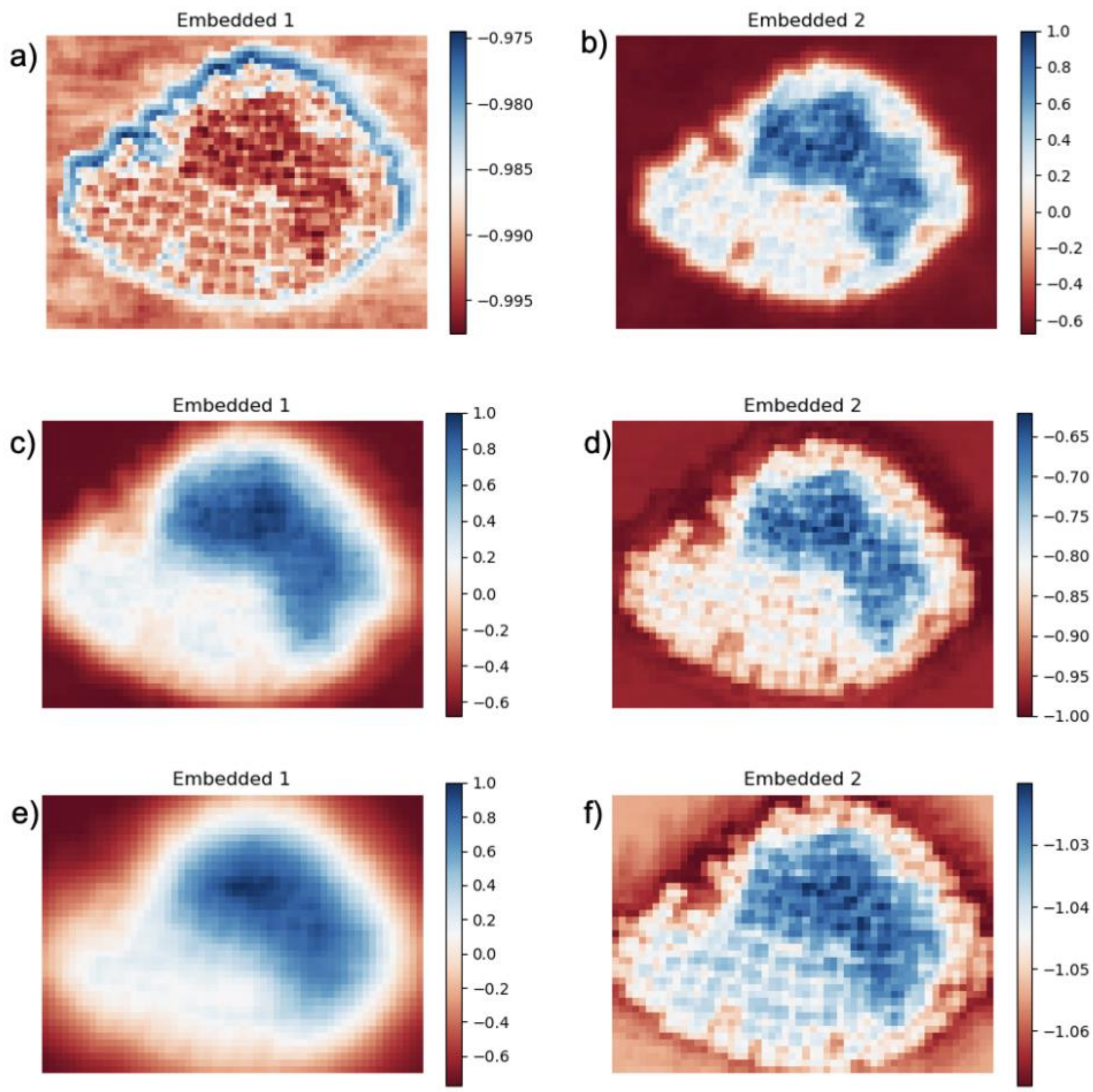

**Figure 4** Showing DKL embedding for scalarizer “3”. a) and b) correspond to patch size 5, c) and d) correspond to patch size 10, e) and f) correspond to patch size 15. See supplementary for all simulation. Here, 1 pixel corresponds to 5.10 nm.

The complementary information can be derived from the classical variational autoencoder (VAE) analysis of the structural data only. Over the last several years, VAEs have emerged as a powerful tool for building low-dimensional representations of data in the form of latent vectors. The encoder part of the VAE compresses data to the latent vector, whereas decoder expands the latent vector back to the original dimensionality, balancing the reconstruction loss and the Kullback-Leibler loss between latent distribution and Gaussian. The key aspect of the VAE is their capability to disentangle the factors of variability in the data, for example the width and tilt of handwritten digits. These can be conveniently represented for the 2D space as latent representations as shown in Figure 5. In this, the 2D latent space of the trained VAE is sampled over 2D grid, and reconstructed objects are plotted as an image. The applications of VAEs for imaging data are discussed in depth in Refs[29–31]. In particular, VAEs also allow to explicit separation of invariances in data, for example rotations or translations. This comes handy in microscopy images as we often visualize lattices with symmetry in both translation and rotation, if the model is unaware about this inductive bias, it will have to see a lot of data to just learn symmetry. Having this already encoded in model helps in better model performance in active learning setup where data is already scarce. The rotationally invariant VAE (rVAE) will discover the features with any rotational angle and separate it as an additional physically defined factor of variation.

Here, we note that the structure of the DKL latent space is determined both by the structural and spectral features. Conversely, the structure of the VAE latent space is determined only by the data itself. Due its capability to disentangle the latent representations, VAE gives us access to a feature space which is helpful in navigating the search space. For example, the initial selection of windows size can be guided by this analysis based on structure of latent representations and complexity of latent distributions. For example, we can see the scalarizer property is highlighted better in embedding of DKL with patch size 5.

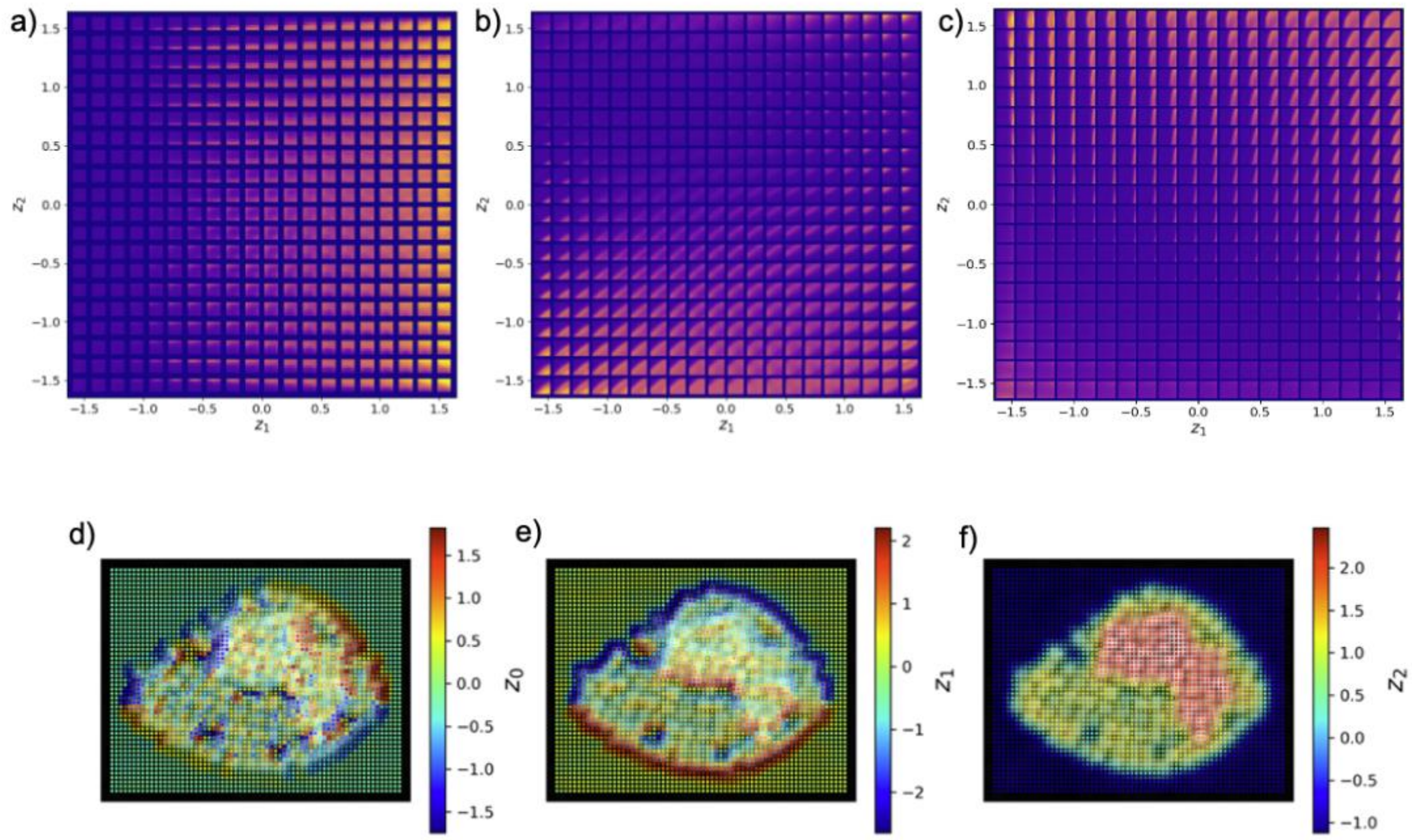

**Figure 5**. a), b) and c) showing latent space of rVAE for patch size 5,10 and 15 respectively. d, e and f are corresponding latent images for patch size 5. Note that the image patch can be decoded from any point in the latent space. The latent representations (top row) are generated via decoding the rectangular sampling grid in the latent space of VAE and illustrate evolution of features over the latent space. Angle latent vector is represented by $z_0$ in radians.

## IV. DKL active learning

In DKL active learning, ML agent issues the commands to the microscope. Human operator can amend the ML behavior via choice of policies and scalarizer. However, the steering of the AE requires the monitoring the progression of the DKL experiments. Here we explore these monitoring functions and show how hyperparameters of DKL algorithm affect the process.

In the actual automated experiment, we always must contend with drift, beam damage, and other non-stationary effects. In order to simulate a wide range of scenarios, stress-test our system under various conditions, and fine-tune the active learning algorithms to achieve optimal performance here we explore the AE using the pre-acquired data.

### IV.1. Monitoring learning

The first set of monitoring variables are directly available from the DKL itself, namely the predicted scalarizer and predicted uncertainty. Note that by the nature of the DKL experiment, these are defined for all patches within the image. Hence, for prediction and uncertainty we can visualize the overall behavior, including the mean and dispersion, and explore the evolution of the full distribution functions.

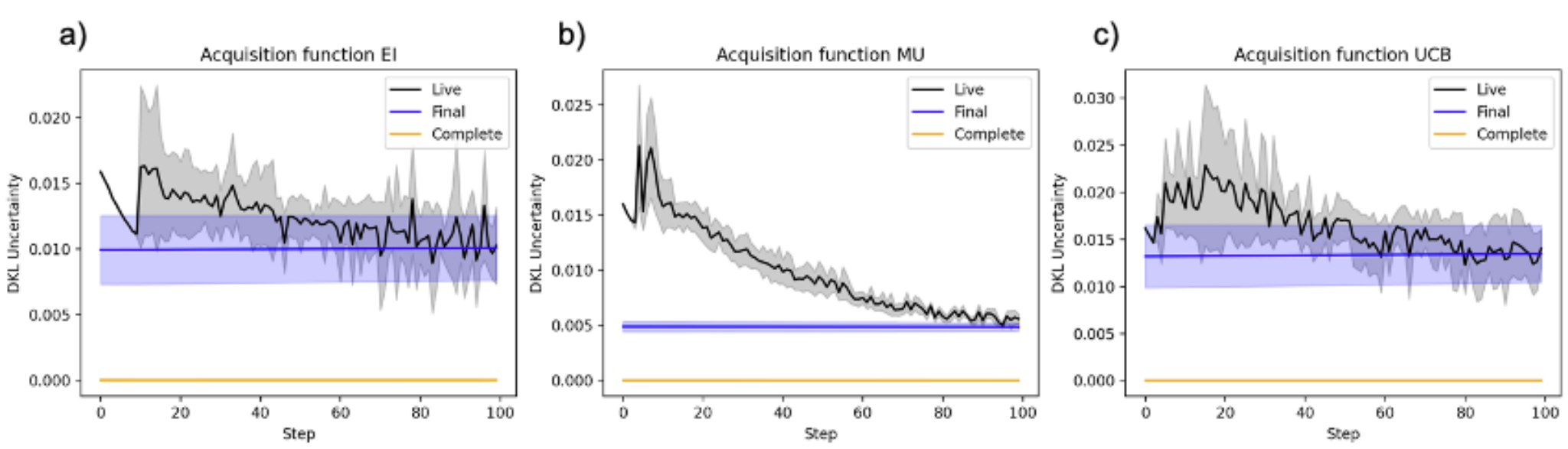

d)

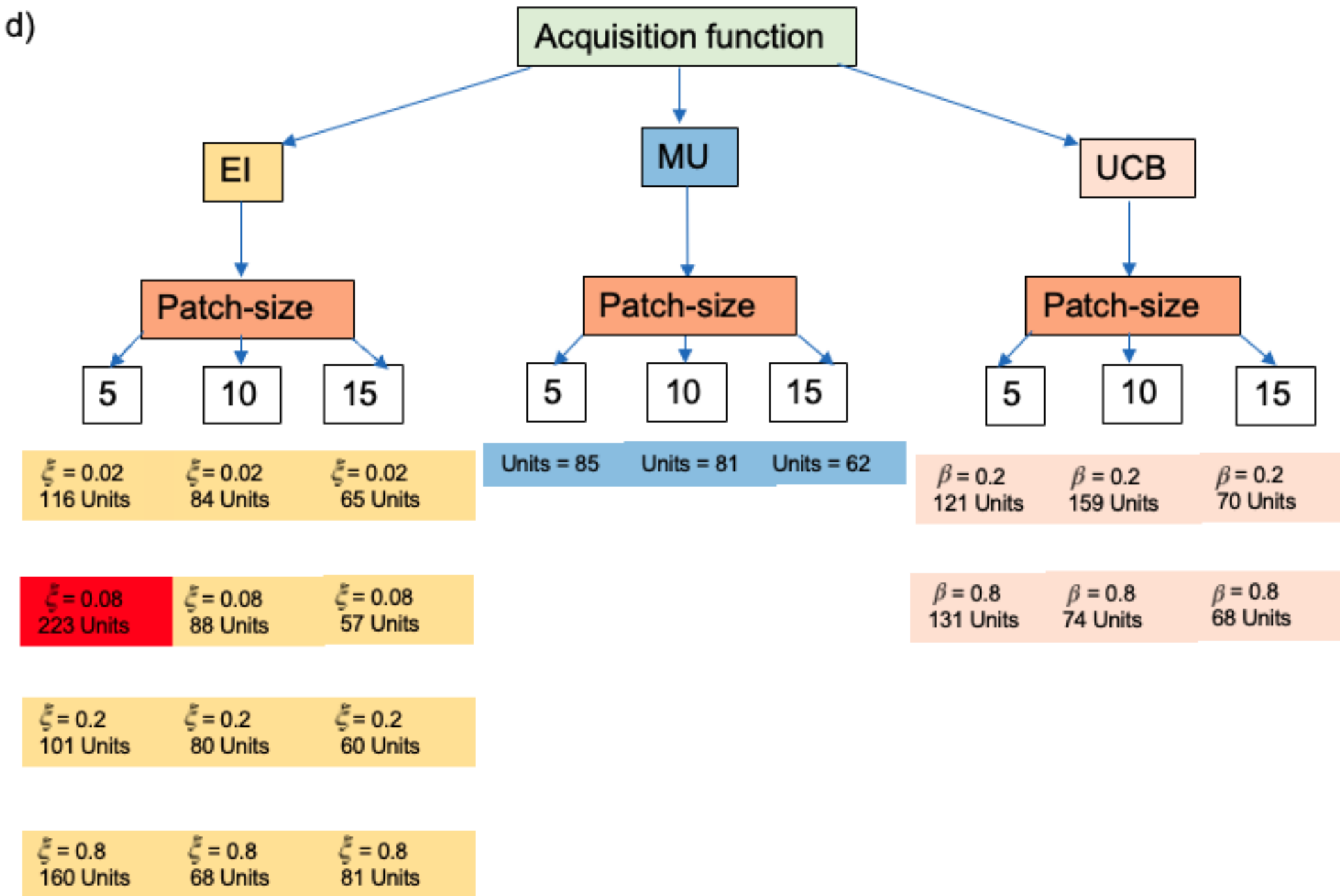

**Figure 6**. Monitoring learning using predictive uncertainty(values multiplied by 10000). a) is for acquisition function EI b) for MU and c) for UCB. Black, blue and orange curve represents live, final and complete model. The predictive uncertainty values varying "$\xi$" and "$\beta$ " values illustrated in d). Note: The values in "d" have been scaled by a factor of 1000.

As an example, shown in Figure 6 are the learning curves for the predictive uncertainty, with bold black line represents mean of the prediction from the model and shaded region correspond to uncertainty intervals. Here, the mean and dispersion of uncertainty are calculated for all structural patches. The mean hence quantifies average uncertainty for prediction of scalarized for all patches. The corresponding dispersion quantifies the distribution of uncertainties over the collection of the patches.

For comparison, we also show the predictions of the final and complete models. The final model coincides with the live model by the end of the experiment, whereas complete model has been trained on the full data (patch-spectrum pairs), and provides the comparison point for the effectiveness of learning. Generally, of interest is the overall learning dynamics, namely the rate of the evolution of the predictive uncertainty and its distribution, and the knowledge gain (decrease of uncertainty) from the initial state and closeness to the predictions of the complete model.

The analysis of the learning curves for multiple scenarios as described in Figure 8, reveals a spectrum of potential behaviors, with detailed variations outlined in the appendix. Depending on the parameterization, the learning progression may exhibit a rapid decline followed by a plateau, an exponential-like decrease as exemplified in Figure 6b or display intermittent jumps indicative of sporadic learning phases. Crucially, the variance in predictive uncertainty serves as a gauge for the stability of the learning process. As evidenced by the comparative analysis between Figures 6a and 6b, it is apparent that the latter demonstrates a more consistent and stable learning trajectory. This stability reflects the reliability of the learning algorithm in developing an accurate model over the course of iterative training sessions. Such insights are invaluable for refining the active learning framework, guiding the selection of parameters that foster a balance between rapid convergence and consistent learning stability.

**IV.2. Monitoring discovery**

The second observable which aids AE is monitoring learning as described by **next step uncertainty**, as shown in figure 7**.** Panel a), b) and c) shows how the uncertainty varies for the next step over the experimental trajectory for different acquisition functions, panel d) further quantifies the values, showing maximum(non desirable) for configuration "$\xi$"=0.08. For the STEM-EELS data explored here, the evolution of the model prediction is typically very noisy. We attribute this behavior to the presence of multiple geometries with almost equivalent values of the scalarizer function, resulting in a very shallow landscape for the acquisition function. This supposition is further confirmed in section below(**IV.3)**.

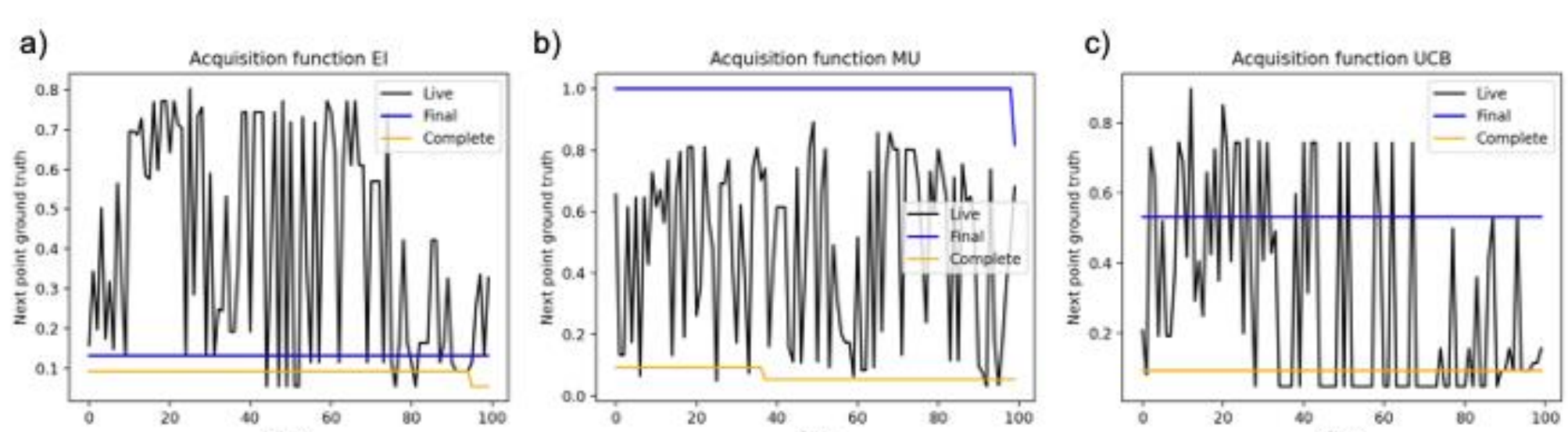

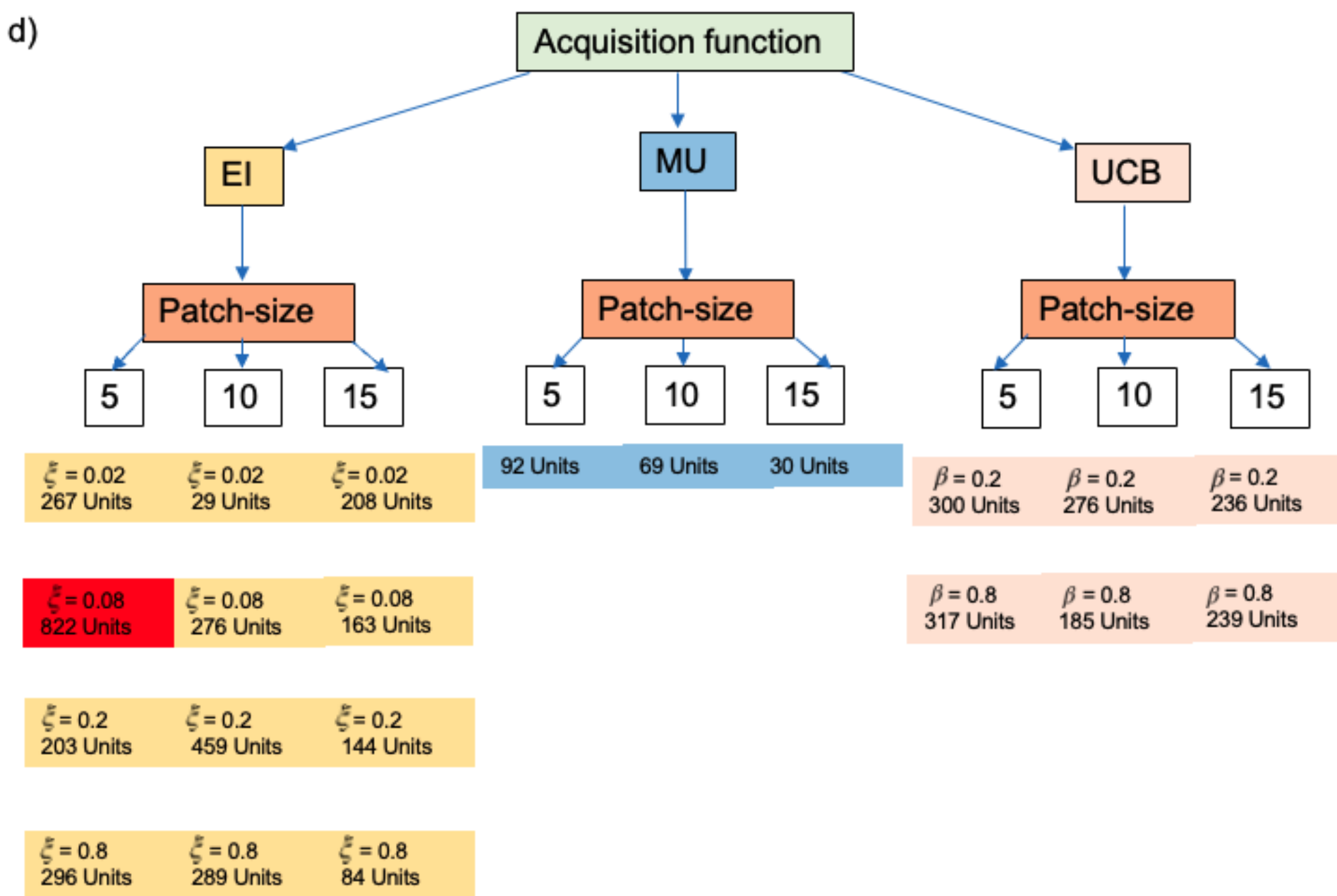

**Figure 7.** a), b) and c) shows next step uncertainty evolution with steps for three acquisition function. The next step uncertainty values varying "$\xi$" and "$\beta$ " values illustrated in d). Note: The next step uncertainty values(represented as units) in "d" have been scaled by a factor of 1000.

### IV.3. Monitoring experimental progression in real space

The third monitoring parameter that readily emerges in the context of the DKL STEM-EELS experiment is the experimental trajectory in real space, i.e. the sequence of measurement points selected by the algorithm. We define distance travelled in trajectory as:

$$\sum_i \left( \sqrt{\left(\left(x_i - x_{i+1}\right)^2 + \left(y_i - y_{i+1}\right)^2\right)} \right) \quad (4)$$

where, i is trajectory point goes from 1 to 100, $x_i$ represents movement in x direction, $y_i$ represents movement in y direction.

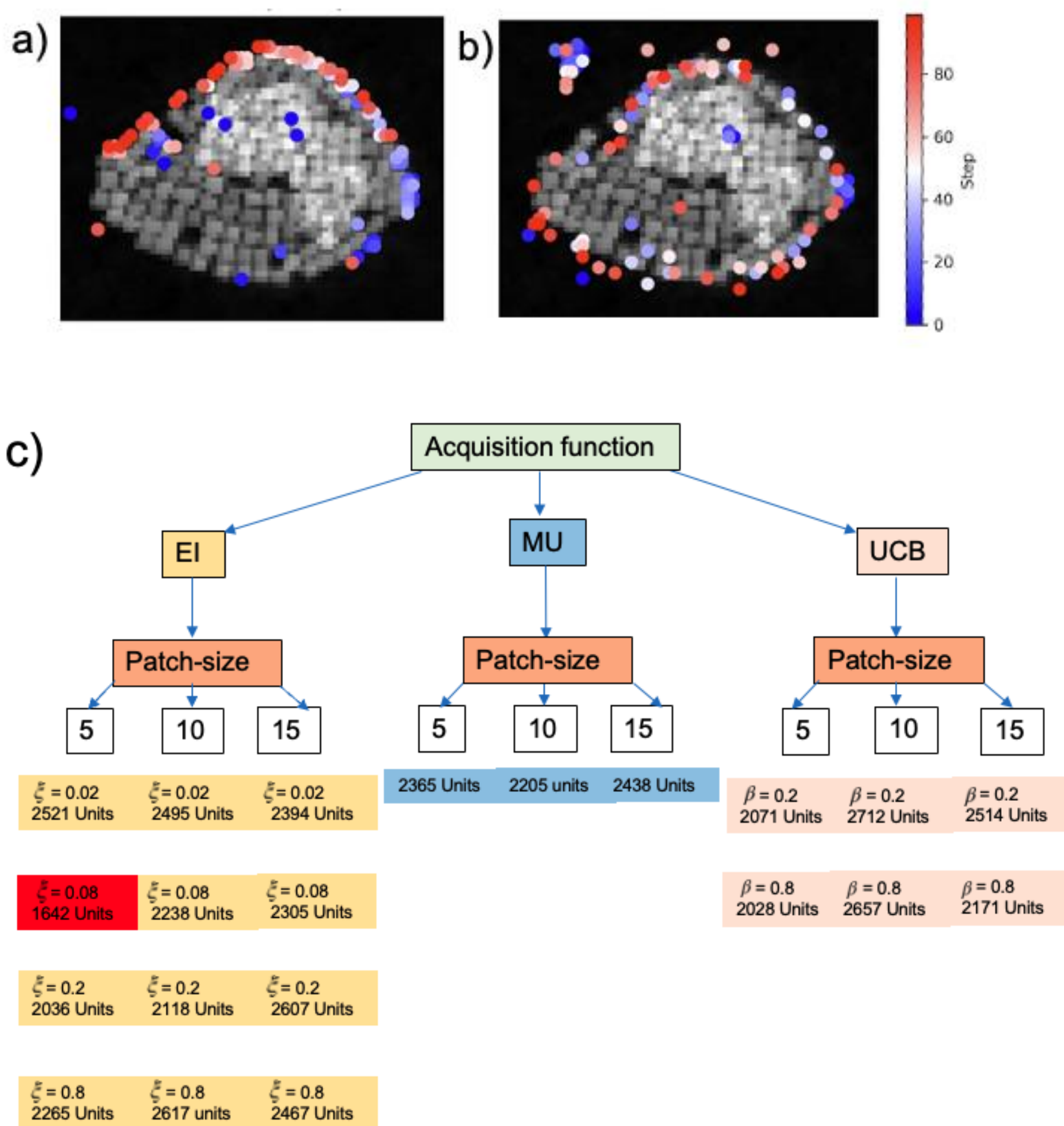

**Figure 8**. a) and b) **s**howing 2 examples of AE experimental progression with acquisition function EI ($\xi$ = 0.08, patch size =5) and EI($\xi$ = 0.02, patch size =10) in the simulation where trajectory exploring the edge and other getting stuck. The trajectory traversed(quantification introduced in eqn 4) values varying "$\xi$" and "$\beta$" values illustrated in panel d).

In comparing the processes described in figures 8 a) and b), we see distinct search behaviors within the parameter space based on the patch size and $\xi$ value. For figure 8 a), a process utilizing a patch size of 5 and an $\xi$ value of 0.08 yielded a travel distance of 1642 units. This suggests a more direct or constrained search path, which indicates a focused exploration around areas of higher expected improvement. Contrastingly, figure 8 b) presents a process with a patch size of 10 and an $\xi$ value of 0.02, resulting in a travel distance of 2495 units. The longer distance and larger patch size suggest a broader, more exploratory search behavior that covers diverse regions within the parameter space, potentially offering an advantage in avoiding local optima.

The examination of active learning trajectories reveals a complex relationship between the learning process and the chosen hyperparameters. For most scenarios, the trajectory starts with active exploration of the image space at the initial stages of the active learning. However, upon exploration the trajectory can get trapped at the specific minimum. This behavior is particularly often for the smaller patch sizes, as demonstrated by a patch N = 5, and Mean Uncertainty (MU) policy. In contrast, larger patch sizes exhibit a lower propensity for the trajectory to become stuck in the local minimum, suggesting a direct link between the patch size and the trajectory's susceptibility to stagnation.

We note that for the explored scenarios as summarized in Figure 8 d), there is no clear correlation between chosen policy, policy hyperparameters, and window sizes that can guarantee the lack of local minima. In principle, one way to address this problem may be via the introduction of additional components to the acquisition functions that de-prioritize the already explored areas. Similarly, the acquisition function can include the cost of measurement, e.g. the time associated for the traversing form one image location to the next one. We expect these additional components to be highly instrument specific and to be optimized for specific instruments. However, from the general perspective these additional policies will further introduce additional hyperparameters, necessitating the development of both monitoring and intervention strategies, as discussed below.

## IV.4. Monitoring in feature space

We define the feature space of the system as the latent representation of the variational autoencoder trained on the full set of patches. This approach allows to use full power of simple, joint, semi-supervised, and conditional autoencoders to identify relevant aspects of materials structure. The detailed discussion of the VAE for materials structure exploration are discussed in multiple previous references[45,46].

We note that the capability of the VAEs to disentangle factors of variation within the data provides a very powerful tool for the exploration of the materials structure visualized via latent reorientations and latent distributions. The addition of the rotation and translation invariances naturally allows to compensate for the uncertainty in the object selection and presence of the rotational disorder in the system. Finally, semi-supervised VAE approach allows to incorporate prior knowledge on objects of interest (e.g. preferred classes), combining the classification and representation disentanglement tasks.

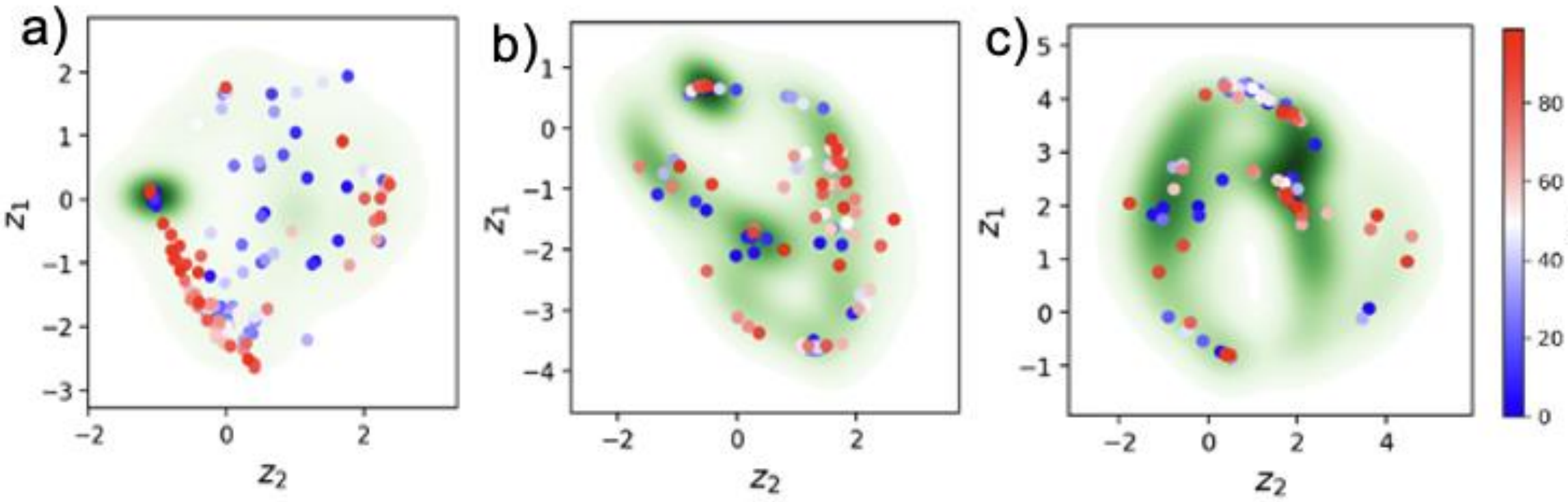

**Figure 9.** a), b) and c) shows example trajectory of experiment progression in rVAE latent space[see Table 1 for definition] for patch size 5, 10 and 15 respectively.

The trajectory of the automated experiment can then be illustrated in the VAE space of the system. Aggregation spots can be observed in latent space of VAE as shown in Figure 9 a), b), c). The latent space of VAE with patch size 5 has one aggregation point, patch size 10 has majorly two aggregation points and several dense regions where experiment progresses. Similarly for path size 15. This behavior is very interesting as it encodes the structural information and at various patch sizes.

**V. Interventions**

The simulation studies above illustrate that the progression automated experiments in STEM-EELS can be monitored based on the learning curves of the DKL model, real space, and feature space trajectories. At the same time, for the certain parameter values the experiment can be trapped in the local minima both in the real and feature spaces. The corresponding behaviors in the parameter space, while demonstrating certain trends, can be highly irregular, necessitating the strategies for real-time interventions during automated experiment. Note that these have to be dynamic almost by definition, given the active nature of real experiment compared to the static nature of the data in classical ML benchmarks (or example workflows used here).

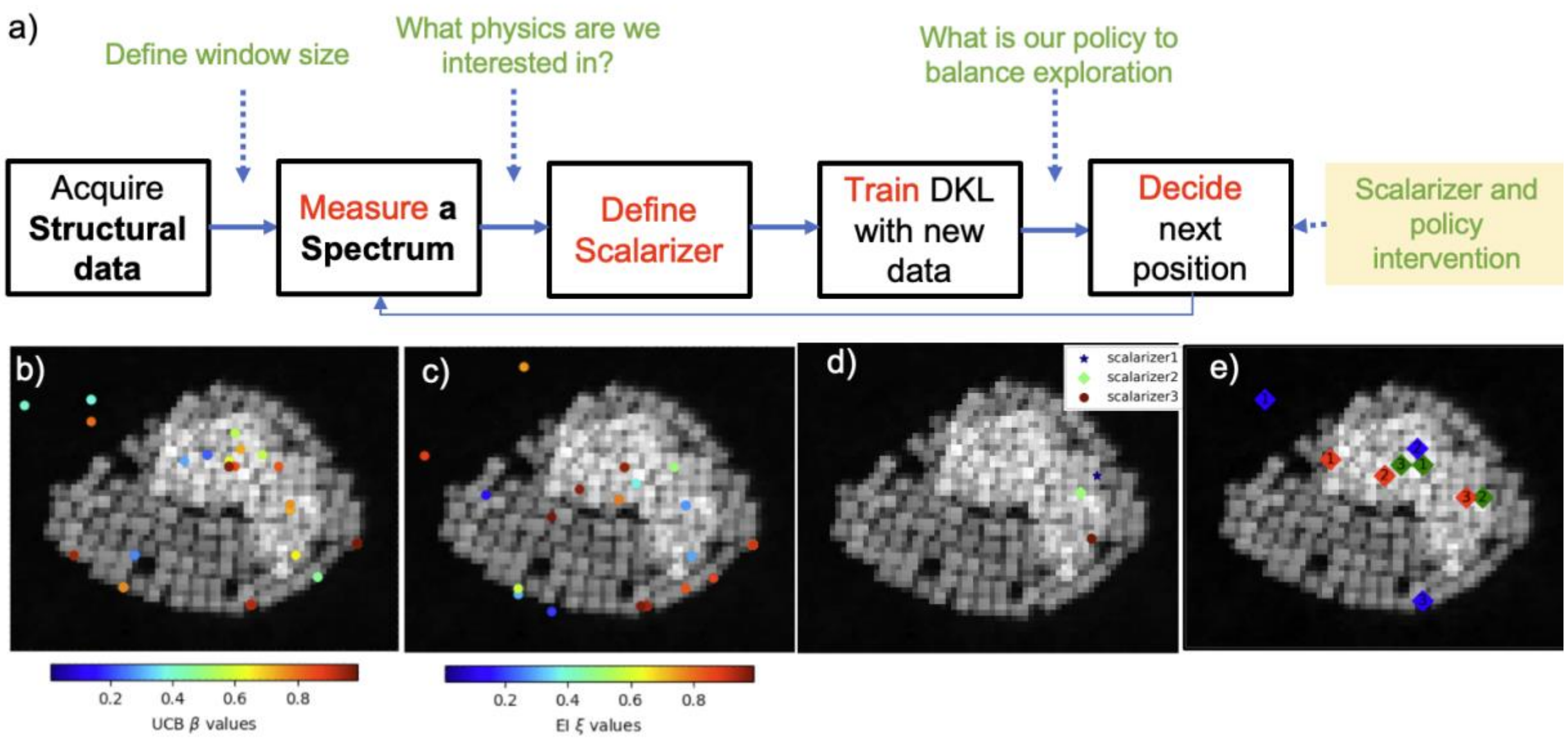

**Figure 10**.a) Shows the overall interactive experimentation flow. b) and c) shows how next point acquisition change with change in policy parameters, called **policy intervention.** Similarly, d) shows scalarization effect for next point acquisition referred to as **scalarizer intervention,** e) demonstrates scalarizer tuning effect for three tuning intervals shown for three steps.

Here, we identify the possible interventions in the DKL workflow. We note that the initial step of the DKL workflow is the selection of the patch size and initial seed points. The effects of the seed points have been explored by Slautin[47] .The patch size effects can be explored prior to the experiment using the VAE feature space exploration, allowing the complexity of the latent distribution and the nature of the disentangle factors of variations to be tuned. We further note that in principle the patch size can be varied during the experiment, i.e. this is a valid intervention. This is turn requires the retraining of the whole model based on the new patch size. It is important to realize that in this case the full experimental trajectory prior to intervention will correspond to off-policy process and can at best be considered as a new extended seed.

The second intervention channel is the exploration target, or scalarizer. This allows to tune the relationship between the full spectrum and the myopic optimization target. The scalarizers can be chosen from multiple classes (e.g. integral intensity, peak ratio, physics base reconstruction), or tuned within class, e.g. change boundaries of the integrated intensity.

The effect of the scalarizer tuning is illustrated in figure 10 e). Here, the scalarizer is integral of the EELS spectrum over an interval [a, b] where a and b can be tuned based on interested physics. We note that the smooth changes in the integration boundaries result in the formation of several distinct clusters of the possible future points. We attribute this behavior to relatively shallow nature of the acquisition function landscape related to highly degenerate relationship between the local geometries and EELS spectra. The flexibility in choice of the interval i.e. a and b values lead in exploring diverse experimental trajectories converging to interesting properties.

Finally, the policies can be tuned on the fly via the selection and hyperparameter tuning of the acquisition functions. Similarly, to scalarize, these can be visualized via the selection of the Upper Confidence Bound (UCB), Expected Improvement (EI), and Maximum Uncertainty (MU), or parameter tuning. (describe how will you visualize it).

Shown in Figure 10 is the effect of $\xi$ value in EI and the $\beta$ value in UCB being adjusted to fine-tune this balance. The scalarizer is also switched to align with the specific physics of interest, such as interface, bulk, or surface plasmons, as identified by the human expert seen in d).

## VII. Summary

To summarize, here we introduce the detailed framework for the human in the loop automated experiment in STEM-EELS based on the myopic optimization workflows. We describe the intrinsic assumptions of the myopic workflows and illustrate how it can be applied to the active experiment in STEM. Based on the exploration of the broad parameter space of the system for the pre-acquired data, including patch sizes, policies, and scalarizers, we demonstrate that for many parameter combinations that AE can be trapped in local minima. The extensive studies of the hyperparameter behaviors demonstrate that this behavior can be very local, and it is not possible to find universally good and robust hyperparameter values.

We hence introduce the strategies of the interactive automated experiment, in which ML agent issues the control signals to the microscope and the human operator monitors the progression of automate experiment of the suitable time scales. To enable hAE STEM-EELS, we introduce a set of monitoring functions based on the DKL model performance and real-space and feature space exploration. We further discuss the strategies for initial parameter selection based on VAE representations and seed point selection.

We introduce the intervention strategies for the DKL workflows based on object selection, scalarizer tuning, and policy tuning. These strategies have been both operationalized and tested on pre-acquired data and indicate strong degeneracies in the STEM-EELS data sets. We note that while all interventions bring the experiment off policy, this allows the dynamic interaction between the human operator and the microscopes.

Finally, we note that proposed human in the loop approach will be applicable to all other myopic workflows, as long an enabling algorithm can yield predictions of function and uncertainty. This includes those based on ensembled neural networks and physics-informed neural networks, contextual bandits, and many other model classes. Similarly, these workflows can be directly translated to other experimental tools including scanning probe microscopy, chemical imaging, and combinations such as nanoindentation with optical and scanning electron microscopy. As such, these developments are universal and can improve multiple areas of materials science and chemical and physical imaging.

## VIII. Acknowledgements

The development of the control and benchmarking frameworks (UP) was supported by the AI4Tennessee initiative. The development of hAE workflow was supported (SVK) as part of the center for 3D Ferroelectric Microelectronics (3DFeM), an Energy Frontier Research Center funded by the U.S. Department of Energy (DOE), Office of Science, Basic Energy Sciences under Award Number DE-SC0021118. The EELS data acquisition and initial analysis (Y.L., K.M.R.) was supported by the Center for Nanophase Materials Sciences (CNMS), which is a US Department of Energy, Office of Science User Facility at Oak Ridge National Laboratory. The authors acknowledge support from the Center for Nanophase Materials Sciences (CNMS) user facility, project CNMS2023-B-02177, which is a US Department of Energy, Office of Science User Facility at Oak Ridge National Laboratory. This work (GD) was supported by the U.S. Department of Energy, Office of Science, Basic Energy Sciences, Materials Sciences and Engineering Division

**Table 1:**

| Term | Definition | Availability |
|---|---|---|
| Global structural image S(x,y) | Initial dataset of structural information was provided prior to the DKL (Deep Kernel Learning) experiment, which was utilized for generating patches to train the DKL model. | Before |
| Spectrum A(E) | The EELS measurement | During |
| Hyperspectral image A(x,y,E) | Collection of patches and spectra | During |
| Scalarizer function, P | Extract interested physics from spectrum | Before |
| Experimental trace | Spectrum and patches together | After |
| Acquisition function | Decides exploration or exploitation. | Before |
| Policy | The guiding criterion for choosing the next path in the sequence involves, at its most basic, the maximization of the acquisition function. | |
| Live model | The model being trained during the experiment | During |
| Final model | The model as soon as active learning terminates | After |
| Complete model | The DKL model trained from full dataset generated on grid. | NA |
| VAE latent | Trained on full patches | Before |
| Latent space | The lower-dimensional encoded space pertaining to VAE/DKL | During |
| Latent embeddings | The feature representation of input in the latent space | During |
| Latent Images | Representation of entire dataset along each of the dimensions. Number of Latent images = Number of dimensions of latent space | NA |
| Full DKL | Trained on complete data | NA |
| Learning curve | Curve showing how the DKL model behaving in active learning. | During |
| Monitoring curve | Curve representing next point uncertainty | During |

**Supplementary Materials**

| Window size | Scalarizer 1 roughness | Scalrizer 2 roughness | Scalarizer 3 roughness | Average Roughness |
|---|---|---|---|---|
| 5 | 0.32 | 0.36 | 0.36 | 0.346 |
| 10 | 0.35 | 0.32 | 0.31 | 0.326 |
| 15 | 0.29 | 0.27 | 0.30 | 0.286 |

| Window size | Chosen scalarizer | Embedding in full DKL |
|---|---|---|
| 5 | 1 | 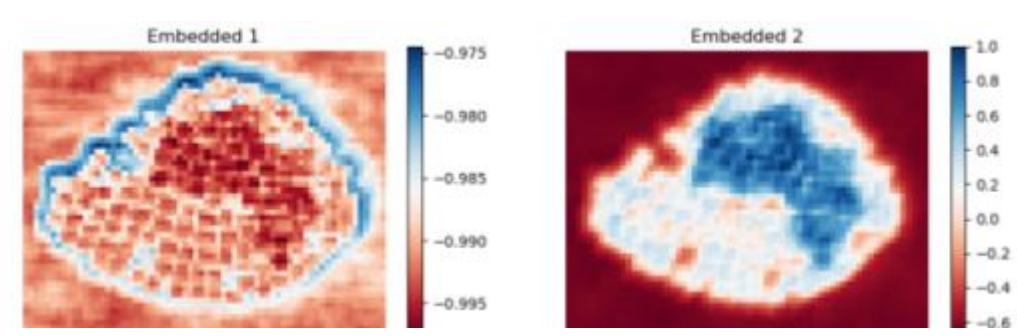  |
| 10 | 1 | 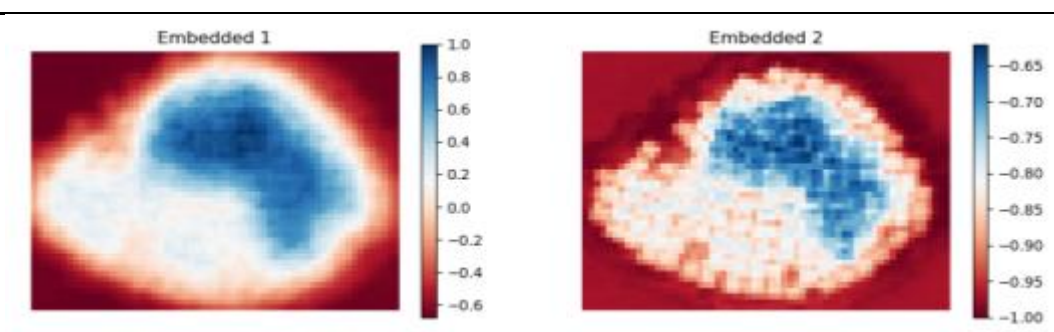  |
| 15 | 1 | 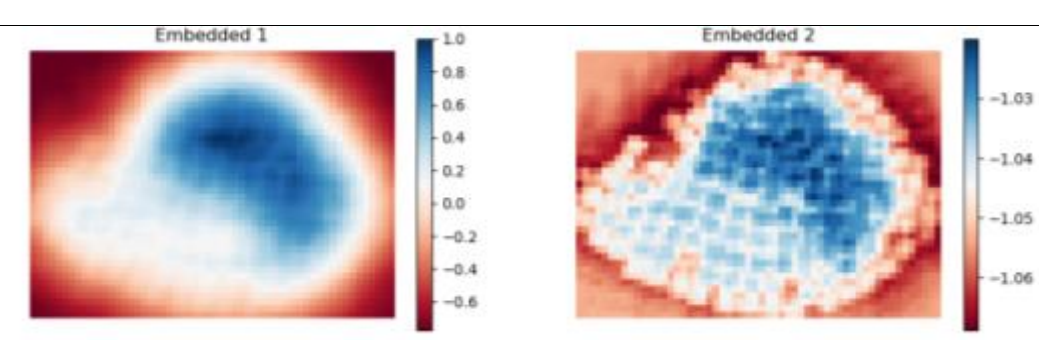  |

| Window size | Chosen scalarizer | Embedding in full DKL |
|---|---|---|
| 5 | 2 | 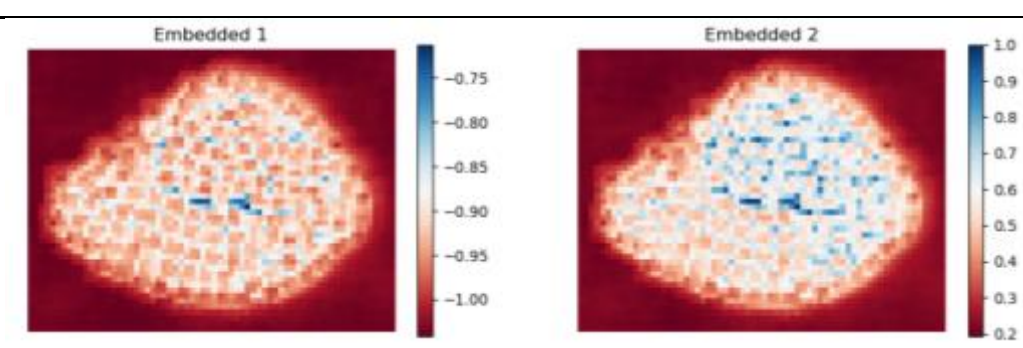  |
| 10 | 2 | 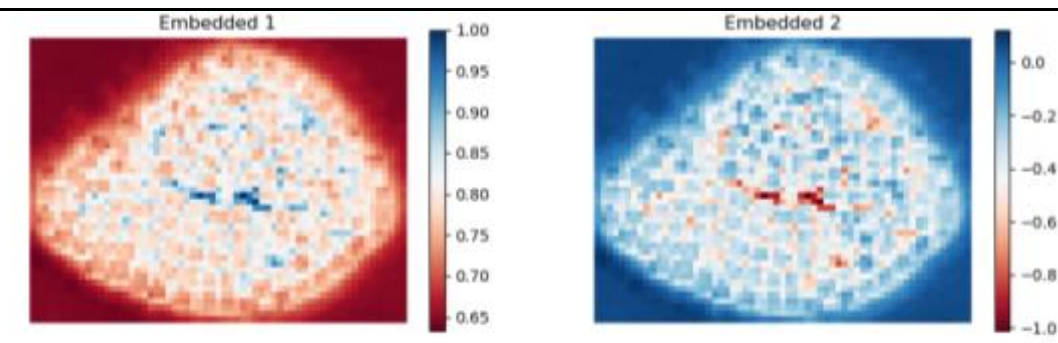 |
| 15 | 2 | 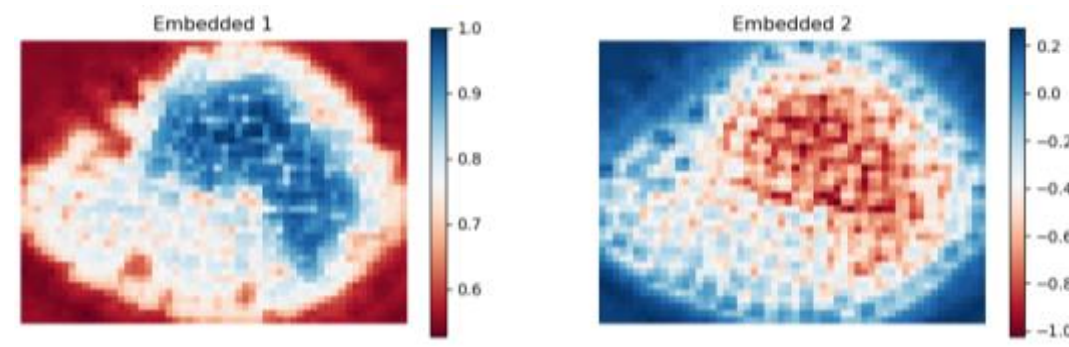 |

| Window size | Chosen scalarizer | Embedding in full DKL |
|---|---|---|
| 5 | 3 | 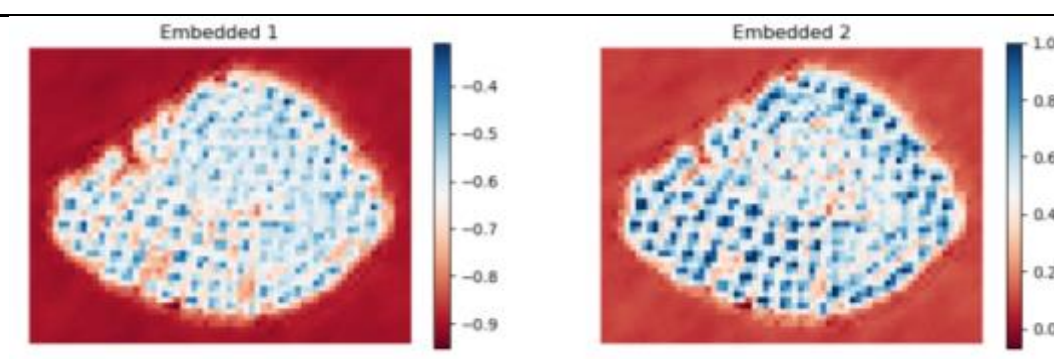 |
| 10 | 3 | 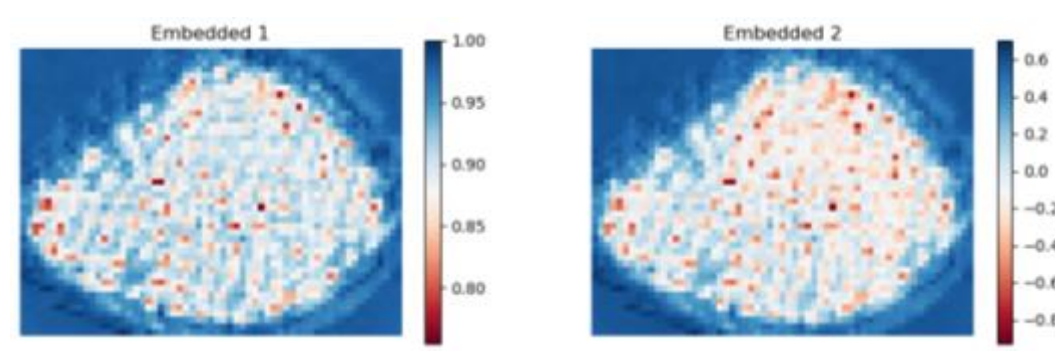 |